# Observation of optical-fiber Kerr nonlinearity at the single-photon level

**Nobuyuki Matsuda[1,2,¶], Ryosuke Shimizu[2,3,†], Yasuyoshi Mitsumori[1,2], Hideo Kosaka[1,2], Keiichi Edamatsu[1,2]**

[1]Research Institute of Electrical Communication, Tohoku University, Sendai 980-8577, Japan

[2]CREST, Japan Science and Technology Agency, Kawaguchi, Saitama 330-0012, Japan

[3]PRESTO, Japan Science and Technology Agency, Kawaguchi, Saitama 330-0012, Japan

[¶]Present address: NTT Basic Research Laboratories, NTT Corporation, Atsugi, Kanagawa 243-0198, Japan; Nanophotonics Center, NTT Corporation, Atsugi, Kanagawa 243-0198, Japan

[†]Present address: Center for Frontier Science and Engineering, University of Electro-Communications, Tokyo 182-8585, Japan

**Optical fibers have been enabling numerous distinguished applications involving the operation and generation of light, such as soliton transmission[1], light amplification[2], all-optical switching[3] and supercontinuum generation[4]. The active function of optical fibers in the quantum regime is expected to be applicable to ultralow-power all-optical signal processing[5] and quantum information processing[6]. Here we demonstrate the first experimental observation of optical nonlinearity at the single-photon level in an optical fiber. Taking advantage of large nonlinearity and managed dispersion of a photonic crystal fiber[7,8], we have successfully measured very small ($10^{-7} \sim 10^{-8}$) conditional phase shifts induced by weak coherent pulses that contain one or less than one photon per pulse on average. In spite of its tininess, the phase shift was measurable using much (~$10^6$ times) stronger coherent probe pulses than the pump pulses. We discuss the feasibility of quantum information processing using optical fibers, taking into account the observed Kerr nonlinearity accompanied by ultrafast response time and low induced loss.**

A photon, the light quantum having much less interaction with its environment than do other quanta (*e.g.*, electron spin, superconducting current), is an outstanding carrier of information for quantum communication and thus is called a 'flying qubit.' This also means that photons may not be suited for computations that require strong unitary interaction between qubits. Hence, fabrication of optical nonlinear media that intermediate sufficiently strong interaction between photons has been under intense study. Cavity quantum electrodynamics-based devices have performed nonlinear Kerr phase shifts of a few ten degrees at the single-photon level[9,10]. Another approach to quantum-optical information processing (QOIP) is to apply *weak* nonlinearity inherent in currently existing media. Recent proposals[11,12] indicated that such moderately weak nonlinearity can mediate the interaction between photons or other qubits through a strong coherent light (known as a qubus). Exploring the availability of single-photon-level nonlinearity in various media is thus an important challenge that provides a test bed for nonlinear optical phenomena that may emerge in the quantum regime of light[13].

In the present experiment, we used a photonic crystal fiber (PCF) as a Kerr medium. PCF has a high capacity for confining light in its silica core by a large core-cladding index contrast[7,8]. Taking advantage of this feature incorporated with its controlled dispersion property, PCF is widely applied in various applications such as supercontinuum generation[4], entangled photon generation[14], squeezing light[15] and a test of the event horizon[16]. To measure the expected ultrasmall phase shift at the single-photon level, we adopted a polarization-division Sagnac interferometer (SI)[17]. The SI has the advantage of inherent stability; two interfering beams counter-propagate through the same path in the interferometer so that the unnecessary reciprocal phase changes originated from the environment are canceled out. Measurement of the ultrasmall phase shifts is a fascinating challenge in metrology, *e.g.*, sensing a molecule in condensed matter[18] and observing gravitational waves from outer space[19].

A schematic illustration of the measurement setup is shown in Fig. 1. The clockwise (CW) probe pulses in the SI experience nonlinear phase shift $\phi_{NL}$ via the cross-phase modulation (XPM) induced by pump pulses, which co-propagate with the CW probe pulses in the PCF. The counter-clockwise (CCW) probe pulses in the SI are used as the reference light of the interference for the phase measurement. In the regime where nonlinear phase shift $\phi_{NL} > 10^{-2}$, we measure $\phi_{NL}$ by a direct projection measurement of the probe-polarization state[20]. In the small-phase-shift regime, a polarization-sensitive optical bridge technique (OBT)[21], which in principle provides shot-noise limited measurements of the phase shift, is used to measure the tiny phase shift. The differential optical power $\Delta P$ at a balanced receiver for the OBT is connected to $\phi_{NL}$ as $\Delta P \approx \phi_{NL} P_{int.}$ when $\phi_{NL} \ll 1$, where $P_{int.}$ is the total interfering probe power of the CW and CCW probe pulses. Therefore, the use of strong probe power enhances the output signal. Details of the measurement methods incorporating further noise reduction techniques are described in METHODS.

Figure 2 shows the measured phase shift $\phi_{NL}$ as a function of average photon number per pulse $\bar{n}_p$ or average pump power $P_p = 2\pi R \hbar \lambda \bar{n}_p$, where the laser repetition rate $R$ was 1 GHz and the center wavelength of the laser $\lambda$ was tuned to 802 nm, which corresponds to the zero-dispersion wavelength of the PCF. Filled and open circles are the data measured by the polarization-projection measurement and the OBT, respectively. The lock-in voltage obtained by the OBT was calibrated to $\phi_{NL}$ so that the data match with those obtained by the projection measurement. The intensity of the pump pulses was controlled by a variable neutral density filter (VNDF) inserted in the pump path. The figure shows clearly that $\phi_{NL}$ varies in proportion to $\bar{n}_p$ in the whole range of our measurement, as expected. We have successfully measured the single-photon phase shift of $9.8 \times 10^{-8}$ with a lock-in time constant $\tau = 3$ s. In addition, $\tau = 30$ s allowed us to measure one order of magnitude less phase shift of $\sim 10^{-8}$ at $\bar{n}_p = 0.1$. Note that this experiment was carried out at room temperature. The probe phase noise levels $\phi_{noise}$ for $\tau = 3$ and 30 s is also shown in Fig. 2. $\phi_{noise}$ was measured as the standard deviation of the OBT output without pump pulses. The origin of $\phi_{noise}$ is discussed later. Evaluation of the observed phase shift and measured dispersion properties of the PCF are available in Supplementary information.

In general, according to the Kramers–Kronig relation, XPM is accompanied by the change in absorption coefficient induced by the pump pulses, namely cross-absorption modulation (XAM). The XAM that is measurable without a quarter-wave plate (QWP) in Fig. 1 is shown by squares in Fig. 2. As the figure shows, the XAM is approximately two degrees of magnitude less than the XPM. This is a characteristic of off-resonant excitation of the silica. In the near-resonant regime, accompanying XAM becomes comparable to XPM[9]. The negligibly small contribution of the XAM in our case is advantageous for the implementation of the quantum gates, because the loss of light inevitably decreases the operation's success probability[13].

Our pump-probe measurement, in which much stronger probe power than that of the pump is inevitably required to obtain high phase resolution, is counter-intuitive, because the authors of experiments on nonlinear photophysics usually set the probe power as weak as possible so that the probe itself does not disturb the result caused by the pump. To show that the variation of $P_{int}$ has no contribution to $\phi_{NL}$, $\phi_{NL}$ is plotted as a function of $P_{int.}$ for several $\bar{n}_p$ in Fig. 3. The figure shows that $\phi_{NL}$ is constant against the variation in $P_{int}$. Therefore, $\phi_{NL}$ depends only on the pump pulses in the range of our experiment, which implies the feasibility of the bus-mediated quantum computation protocol[11,12], in which a strong coherent state undergoes weak Kerr interactions with independent qubits to intermediate their interaction.

It is worth discussing the probe phase noise levels, $\phi_{noise}$, in our experiment. In Fig. 3, the probe phase noise $\phi_{noise}$ at $\tau = 3$ s are plotted under several conditions, as follows: (i) probe phase noise with the PCF, (ii) probe phase noise without the PCF and (iii) total electrical noise of the system. The phase noise consists of the electrical noise, the probe shot noise and the constant phase fluctuation between CW and CCW probe pulses imposed in the PCF. The constant electrical noise $V_{noise}$ turns to the equivalent noise of $\phi_{NL}$ that is proportional to $P_{int.}^{-1}$, because the signal $V$ is proportional to $\phi_{NL}P_{int.}$. The probe shot noise and the constant phase noise have $P_{int.}^{-0.5}$ and $P_{int.}^{0}$ dependence, respectively. Thus each set of data could be fitted by $\sigma P_{int.}^{-1} + \mu P_{int.}^{-0.5} + \nu P_{int.}^{0}$. The dot-dashed curve (i) was fitted by this function. For curves (ii) and (iii) we used the same parameters as obtained from (i), but we set $\nu = 0$ for (ii) and $\mu = \nu = 0$ for (iii). From the data (i), it can be seen that probe phase noise decreases with an increase of $P_{int.}$ and then bottoms out at $\nu$. From the fitting of (i) we obtained $\nu$ as $2.0 \times 10^{-8}$, which is in between the probe shot noise level and the single-photon phase shift at around $P_{int.} = 1$ mW. This low $\nu$ was achieved by the use of Faraday units (FUs), which make CW and CCW probes have the same linearly-polarized mode in the PCF so that temporal phase fluctuation induced by changes in fiber birefringence is minimized. In fact, $\nu$ becomes approximately three times larger when the counter-propagating probe modes have different linear polarizations, as indicated in Fig. 3 (iv); in this situation it becomes more difficult to measure the single-photon phase shift. In curve (i), $\phi_{noise}$ becomes constant with the increase of $P_{int.}$. However, the contribution of the self-phase modulation (SPM) in the PCF would be dominant with increasing probe power. The curve (v) is the estimated probe phase noise taking SPM into consideration[22]. The SPM results in squeezing and antisqueezing in phase space[23] and thus possibly limits the sensitivity of the phase measurement. From curve (v), we see that the SPM effect becomes dominant when $P_{int.}$ is larger than a few mW and that the probe power we used (around 1 mW) is in the range of minimum $\phi_{noise}$. In addition, the insensitivity of birefringence against the temperature fluctuation and reduced guided acoustic wave Brillouin scattering in the PCF[24,25] may have contributed to the exceptionally low phase in our experiment.

In Fig. 4(a), we show the pump-delay dependence of the normalized phase shift, $\phi_{NL}/\bar{n}_p$, at $\bar{n}_p \sim 1$, $10^2$ and $10^4$. The full width at half-maximum (FWHM) of approximately 0.6 ps, which corresponds to the autocorrelation width of the pulsed laser, is maintained irrespective of the pump power, which means that the instantaneous XPM response is unchanged regardless of the pump photon number until the single-photon regime. The ultrafast relaxation time (of the order of fs) of nonlinear dielectric polarization is attributed to the off-resonant electronic process in glasses[26]. The result shows the availability of high-speed QOIP using a PCF.

We have experimentally demonstrated that the nonlinear phase shift ($\sim 10^{-8}$) induced by a single-photon-level optical pulse in an optical fiber is measurable. To the best of our knowledge, this is the first experimental demonstration of the single-photon-level nonlinear phase shift in an optical fiber. In our demonstration we used repeated measurements incorporating strong probe pulses ($\sim 3 \times 10^6$ photons per pulse), higher repetition rate (1 GHz) and lock-in detection (time constant = 3 s); the equivalent probe photon number for the measurement is approximately $10^{16}$, and the resulting shot-noise limit of the phase measurement is $\sim 10^{-8}$, as shown in Fig. 3. On the other hand, the implementation of the QOIP requires the probe phase shift that is larger than the phase noise inherent in every single probe pulse. In our experiment, the optimum probe power was $\sim 10^7$ photons per pulse, since higher power would result in a rapid increase of the phase noise, as indicated by the curve (v) in Fig. 3. As a consequence, the measurable phase shift in the single-shot measurement is $\phi_{NL}$ of $\sim 10^{-4}$. This value would be obtainable if optical fibers a few kilometers long with the same nonlinearity as ours combined with reduced loss of 1 dB/km and flattened group-velocity dispersion (GVD) were available. Those demands were recently realized independently[27,28]. However, the increase of fiber length (or nonlinearity) at the same time increases the phase noise originating from the antisqueezing, so it may remain difficult to realize the QOIP, including Munro and Spiller's approach[11,12], in the extrapolation of current fiber technology. However, the convenience of our experimental method, including its room-temperature operation, short measurement time, low interaction loss and tolerability against the use of a strong coherent probe, has not been realized in other experiments on single-photon-level Kerr nonlinearity[9,10]. The technology developed in the present study will be a useful tool for not only proof-of-principle experiments of QOIP but also further advanced studies of quantum nonlinear phenomena not yet revealed.

# METHODS

### Experimental detail

The output spectrum of laser pulses from a mode-locked 1-GHz Ti:Sapphire laser, GigaJet 20C (Giga Optics, GmbH), with 18 nm full-width half-maximum (FWHM) is filtered by an additional grating, a cylindrical lens and a slit narrowed to 4.5 nm. The spatial mode is also cleaned by a subsequent pinhole with a pair of focusing lens inserted before half-wave plate (HWP) 1, so that the coupling efficiency of the incident beam onto the PCF is improved. The optical pulses are divided into pump and probe pulses by a polarization beam splitter (PBS) 1. By HWP 2, the probe pulses are set to have 45° linear polarization, *i.e.*, 1:1 superposition of horizontally (H) and vertically (V) polarized components that counter-propagate through a polarization SI in which a PCF is set. Two Faraday units (FUs), each of which consists of a 45° Faraday rotator and a half-wave plate, make the two probes have the same linear polarization in the PCF to minimize the relative-phase noise that arises from the temporal fluctuation of birefringence between two major fiber axes. V-polarized pump pulses are synchronized with clockwise (CW) probe pulses, and coupled in and decoupled at PBS3 and 4, respectively. In the PCF, the pump pulses induce nonlinear phase shift $\phi_{NL}$ that is proportional to $P_p$ or $\bar{n}_p$ to the CW probe pulses. Thus, the output probe pulse from the SI has the state of polarization of $\exp(-i\phi_{NL})|H\rangle+|V\rangle$. A subsequent QWP converts the polarization into linear polarization with a direction along $\phi_{NL}/2$, which is to be measured by a polarization-projection measurement. More details about this process are available in Ref. 12.

By a projection measurement using one output port of PBS5, we can obtain a reliable amount of large phase shift free of influence from an imbalance between CW and CCW probe powers originating from, for example, the XAM. On the other hand, the OBT allows us to measure a tiny phase shift using PBS5 and a balanced optical receiver. If the initial polarization angle before the PBS5 is set to 45°, the bridge is balanced. Tiny additional rotations in polarization are sensitively measurable. In addition, we used lock-in detection synchronized with the 26-kHz chopping of the pump pulses by an acousto-optic modulator (AOM) to suppress background noise originating from amplitude noise of the laser, which was present mainly in a frequency range lower than 10 kHz in our experiment.

The PCF we used is commercially available, 5 m long NL-2.4-800 fabricated by Blaze Photonics, Inc. The zero-dispersion wavelength was measured as 802±2 nm, to which central the wavelength of the laser is tuned. The linear attenuation coefficient is 80 dB/km, which corresponds to approximately 9% loss for the length of the PCF. The nonlinear coefficient $\gamma$ = 70 /W km. The incident light is focused to the PCF by a ×60 microscope objective lens with the coupling efficiency of approximately 50%. The incident linear polarizations of the probe and the pump pulses are rotated to align to two orthogonal major axes of the PCF by HWP 3 and 4.

## Noise reduction

A pulsed laser with a high (1 GHz) repetition rate was used to increase the number of pulses to be detected per unit of time. In addition, the peak power of each probe pulse can be reduced while the average probe power is maintained. Therefore, probe-induced nonlinear optical effects that degrade the coherence of the probe pulses are successfully reduced, while the number of detected probe photons that determine the shot noise is maintained.

Approximately 5% of the pump pulses transmit PBS4 and leak to the detector due to random fluctuation of fiber birefringence along the axis. This leaked pump transmission coherently interferes with the probe, and the interference signal is modulated by path-length fluctuation of a Mach-Zehnder (MZ) interferometer, which is used to synchronize the pump and the CW probe pulses. The interference is a significant obstacle when the pump power is weak, because the interference is proportional to $\bar{n}_p^{1/2}$, while $\phi_{NL} \propto \bar{n}_p$. To suppress the noise, we averaged out the path-length difference of the MZ arms by using a 110-MHz carrier frequency shift imposed by the AOM. Since this phase modulation is much faster than the lock-in reference frequency (26 kHz), the interference was successfully averaged out.

**Acknowledgements**

The authors are grateful to Professors K. Koshino and H. Ishihara for fruitful discussions. This research was supported in part by a Grant-in-Aid for Creative Scientific Research (No. 17GS1204) and a Grant-in-Aid for JSPS Fellows (No. 20009351) from the Japan Society for the Promotion of Science.

# Figures

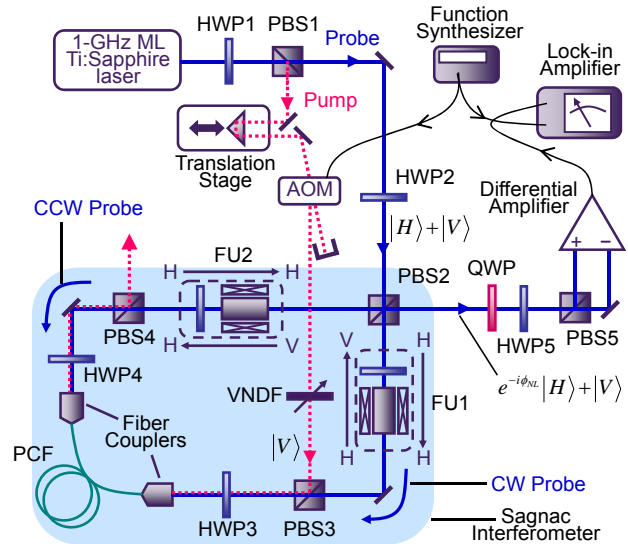

**Figure 1 | Experimental setup.** PBSs, polarizing beam splitters; HWPs, half-wave plates; QWP, quarter-wave plate; AOM, acousto-optic modulator; VNDF, variable neutral density filter; PCF, photonic crystal fiber. FUs are Faraday units, each of which consists of a 45° Faraday rotator and a half-wave plate. $|H\rangle$ and $|V\rangle$ represent horizontal and vertical polarizations, respectively. CW and CCW stand for clockwise and counter-clockwise. The thick lines represent optical connections, and the thin curves are electronic connections.

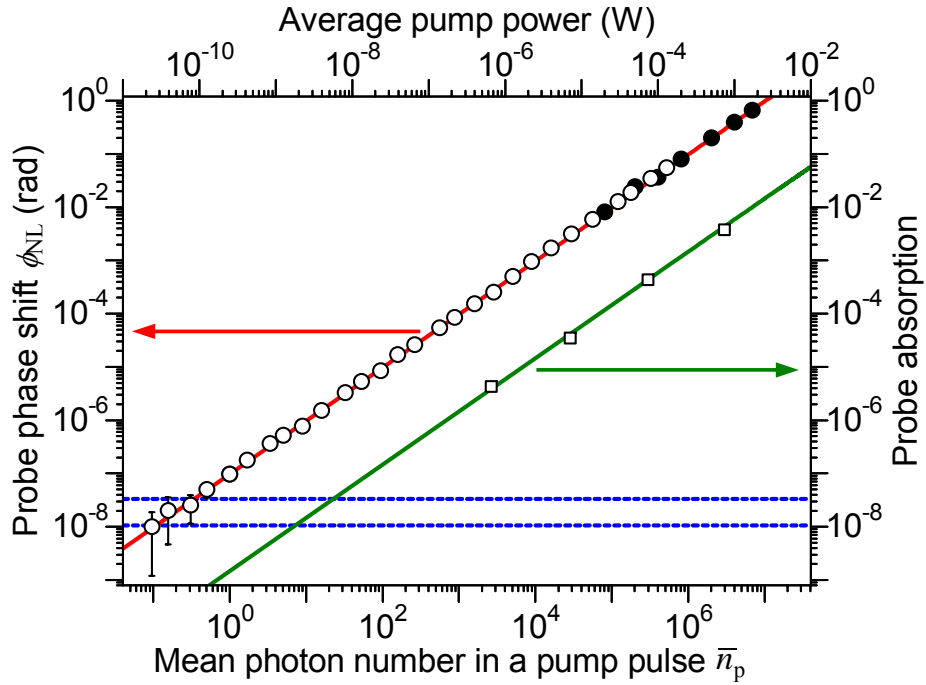

**Figure 2 | Nonlinear phase shifts as a function of mean photon number per pump pulse.** The closed and open circles are the phase shift $\phi_{NL}$ measured by the polarization projection measurement and the optical bridge technique, respectively. All the data for $\bar{n}_p \geq 1$ were measured by the lock-in time constant $\tau$ of 3 s, whereas those for $\bar{n}_p < 1$ were obtained by $\tau = 30$ s. The dashed horizontal lines show the probe phase noise levels for 3 s (upper) and 30 s (lower) lock-in time constants. Pump-induced absorption is also shown by the open squares.

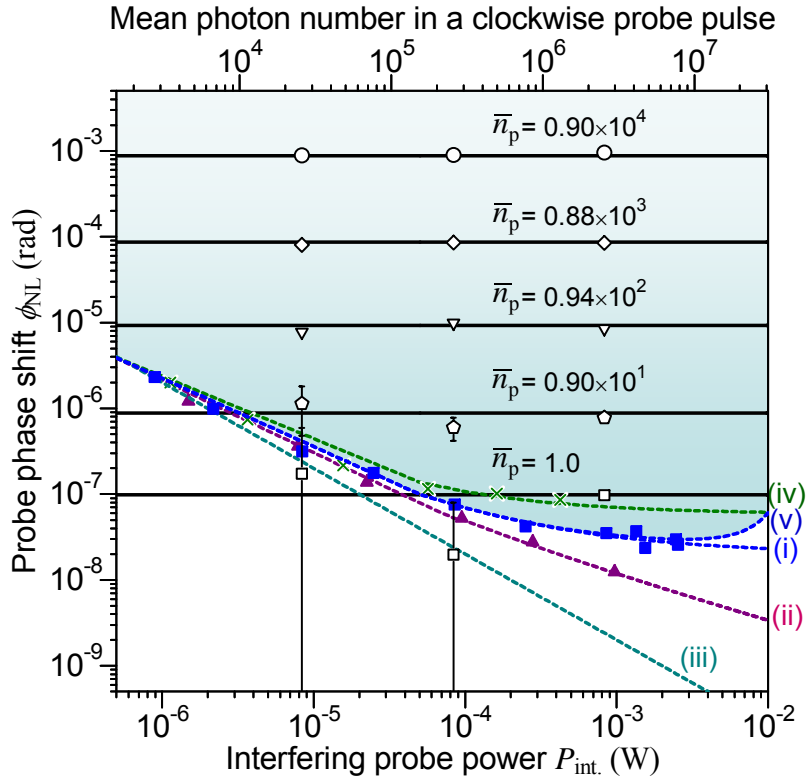

**Figure 3 | Probe-power dependence of the measured phase shift and the phase noise.** The open symbols represent the measured phase shifts for the mean photon numbers in a pump pulse $\bar{n}_p$. The solid-line levels indicate $\phi_{NL}$ obtained from the fitted line in Figure 2. The abscissas are the photon number in a CW probe pulse (top) and the corresponding probe power (bottom). The closed symbols with the curves indicated as (i) to (v) are the noise levels under different conditions described in the text. The phase-shift measurement is accomplished in the hatched region above the noise level. All the data are measured with the lock-in time constant of 3 s.

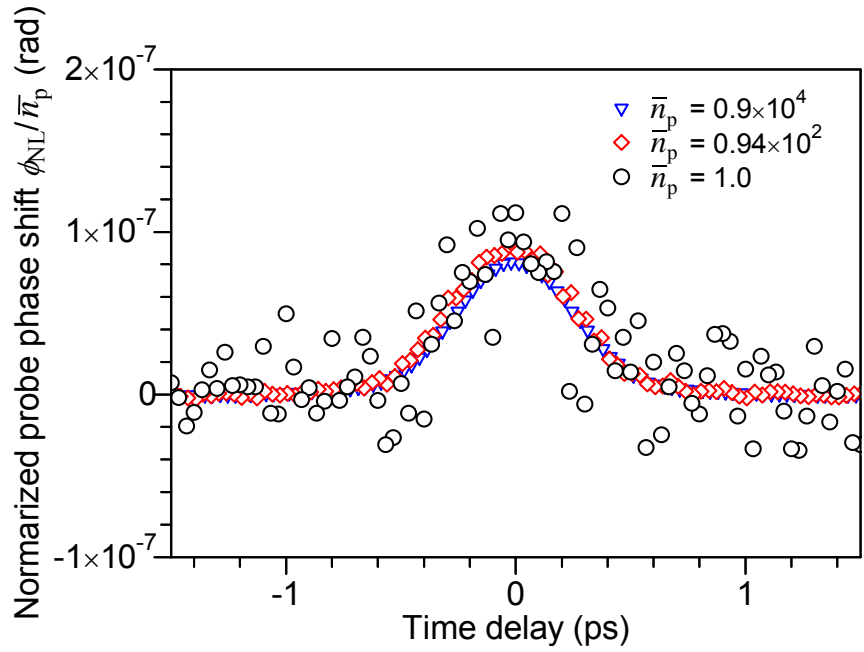

**Figure 4 | Normarized probe phase shifts as a function of the time delay between pump and probe pulses.** The temporal dependence of the phase shift is independent of the pump photon number $\bar{n}_p$ and identical to the laser auto-correlation, indicating the instantaneous response of the induced phase shift.

## Supplementary information

### - Evaluation of the observed nonlinear phase shift

The nonlinear phase shift $\phi_{NL}$ induced by the cross-phase modulation (XPM) is expressed by the following equation:

$$\phi_{NL} = 2b\gamma P L_{eff},$$

where $P$ is the peak pump power, $b$ the polarization parameter associated with the co-propagating modality, $\gamma$ the nonlinear coefficient, and $L_{eff}$ the effective length of the fiber. When the pump and probe pulses are orthogonally polarized, $b = 1/3$. The effective length of the PCF $L_{eff} = (1 - e^{-\alpha L})/\alpha$ was estimated to be 4.5 m with the attenuation coefficient $\alpha = 80$ dB/km and the length $L = 4.7$ m of the PCF. From these values and the slope of the solid line in Fig. 2, $\gamma$ was estimated to be 62 $W^{-1}km^{-1}$, which is in reasonable agreement with the theoretical catalog value (70 $W^{-1}km^{-1}$) of the PCF. The small discrepancy might originate from the decrease in the effective interaction length between the pump and probe pulses caused by polarization-mode dispersion (PMD) in the PCF; the measured value of the PMD was approximately 0.1 ps for the fiber length, which results in the reduction of the interaction length by a few ten percent.

### - Dispersion property of the PCF.

Figure S1 shows the measured group delay as a function of wavelength for the two fundamental polarization modes (A and B; orthogonally polarized with each other) of the PCF. The solid and dashed curves are the second-order polynomial fits to the data. From the fitting, we obtain the zero-dispersion wavelengths and slopes, as summarized in Table S1 with the catalog values.

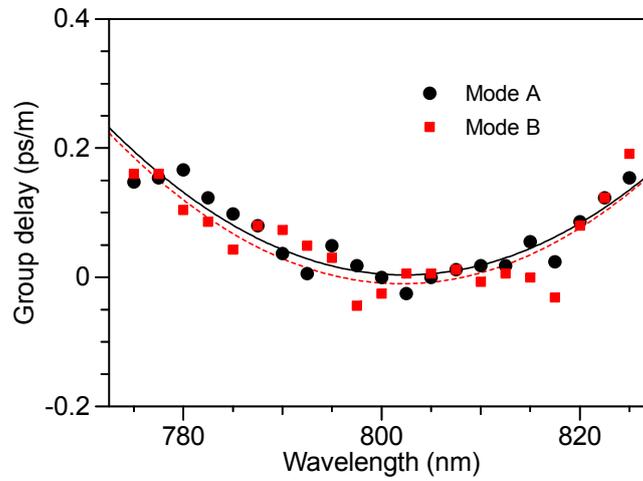

**Figure S1 Measured group delay as a function of wavelength for the two polarization modes of the PCF.**

**Table S1 Dispersion properties of the PCF used in our experiment.**

| Property | Measured value | Catalog value |
|---|---|---|
| Zero-dispersion wavelength | Mode A: 802.3 ± 0.9 nm<br>Mode B: 802.0 ± 0.8 nm | 803 ± 5 nm (average) |
| Zero-dispersion slope | Mode A: 0.61 ps/(nm² km)<br>Mode B: 0.43 ps/(nm² km) | 0.55 ps/(nm² km) (average) |